\documentclass[prl,twocolumn,showpacs,floatfix,amsfonts]{revtex4}
\usepackage{graphicx,graphics,color,epsfig}
\usepackage{bm}
\usepackage{amsmath}
\usepackage{amssymb}
\begin{document}
\preprint{}
\title{
In-plane Tunneling Spectrum into a [110]-Oriented High-$T_c$
Superconductor in the Pseudogap Regime}
\author{Jian-Xin Zhu}
\affiliation{Theoretical Division, MS B262, Los Alamos National
Laboratory, Los Alamos, New Mexico 87545}

\begin{abstract}
{Both the differential tunneling conductance and the surface local
density of states (LDOS) of a [110]-oriented high-temperature
superconductor in the pseudogap (PG) regime are studied
theoretically. As a competing candidate for the mechanism of PG
state, the charge-density wave (CDW), spin-density wave (SDW),
$d$-density wave (DDW), and $d$-wave superconducting (DSC)
orderings show distinct features in the tunneling conductance. For
the CDW, SDW, and DSC orderings, the tunneling conductance
approaches the surface LDOS as the barrier potential is increased.
For the DDW ordering, we show for the first time that there exist
midgap states at the [110] surface, manifesting themselves as a
sharp zero-energy peak in the LDOS, as in the case of DSC
ordering. However, due to the particle-hole pair nature of the DDW
state, these states do not carry current, and consequently the
one-to-one correspondence between the tunneling conductance and
the surface LDOS is absent.}
\end{abstract}
\maketitle

One of the most intriguing properties of high temperature
superconductors is that, in the underdoped regime, the density of
low-lying excitations is suppressed below a characteristic
temperature $T^{*}$, which could be well above the superconducting
transition temperature $T_{c}$~\cite{Timusk99}. The origin for
this so called pseudogap (PG) phenomenon remains unsettled. The
existing scenarios for the mechanism of PG state include: the
preformed pair model, where the Cooper pairs are formed above
$T_c$ but either the phase associated with them fluctuates so
strongly~\cite{Emery95} or the coherence length (essentially the
size of a Cooper pair) is so small~\cite{Chen98} that the
superconducting transition temperature is reduced from $T^{*}$ to
$T_{c}$;
stripes~\cite{Zaanen89,Poilblanc89,Machida89,Emery90,Martin00,Castellani00}
or antiferromagnetic fluctuations~\cite{Pines97}, relying on the
spin and/or charge fluctuations; spin-charge
separation~\cite{Lee99}, where the PG comes from the spinon
sector; quantum criticality~\cite{Sachdev99}, of which the
$d$-density wave (DDW) ordering breaking both time-reversal and
translational symmetry~\cite{Chakravarty01} and the circulating
currents breaking time-reversal but preserving translational
symmetry~\cite{Varma99}, are the most representative models.
Recently, several tunneling experiments have been proposed for
both the preformed pair~\cite{Choi00,Sheehy00} and
antiferromagnetic correlation scenarios~\cite{Bang00}. In this
paper, based on a lattice model, we calculate the tunneling
conductance, within the framework of a scattering approach, of a
[110]-oriented normal-metal/underdoped high-$T_c$ superconductor
junction. We typically consider the possible orderings: (i) the
charge-density wave (CDW) to model the charge stripes; (ii) the
spin-density wave (SDW) to model the spin stripes or
antiferromagnetic correlations; (iii) the DDW state; (iv) the
$d$-wave superconducting (DSC) state. The exposure of distinct
features among these orderings may help us pin down the mechanism
for the PG. On the other hand, it has been predicted~\cite{Hu94}
that a sizable areal density of midgap states (i.e., states with
essentially zero energy relative to the Fermi surface) exists on
the [110]-oriented surface of a $d_{x^{2}-y^{2}}$-wave
superconductor. These states gives rise to a narrow surface local
density of states (LDOS) peak at the Fermi energy, where the bulk
density of states dips to zero. One of the consequences of these
midgap states is a sharp zero bias conductance peak in single
particle tunneling~\cite{Tanaka95,Xu96}, which has been confirmed
by several carefully controlled
experiments~\cite{Covington97,Alff97,Sinha98,Wei98}. A question
arises naturally: Whether such kind of midgap states also exist in
those normal-state orderings, and if so, how to distinguish them
from those in the DSC ordering. This question constitutes another
motivation for the present work. In light of this, we also
calculate the LDOS near the [110]-oriented surface of the
superconductor with all orderings listed above. We choose the
[110] orientation, where the crystalline $a$ axis of the
superconductor has a $\frac{\pi}{4}$ angle with respect to the
direction normal to the surface, because the surface sensitivity
of the electronic structure of some (e.g., the DDW and DSC)
orderings can be manifested most pronouncedly in this setup.

The equation of motion for quasiparticles in a superconductor with
a variety of orderings can be described in a unified manner by the
generalized Bogoliubov-de Gennes equations defined in a
two-dimensional tight-binding  model:
\begin{equation}
\sum_{j} \left(
\begin{array}{cc}
{\mathcal H}_{ij,\sigma} & \Delta_{ij}  \\
\Delta_{ij}^{*} & -{\mathcal H}_{ij,-\sigma}^{*}
\end{array}
\right) \left(
\begin{array}{c}
u_{j,\sigma} \\ v_{j,-\sigma}
\end{array}
\right)
=E
\left(
\begin{array}{c}
u_{i,\sigma} \\ v_{i,-\sigma}
\end{array}
\right)  \;.
\label{EQ:BdG}
\end{equation}
Here  the quasiparticle wavefunction $\left( \begin{array}{c}
u_{i,\sigma}\\ v_{i,-\sigma}
\end{array} \right)$
comprises the electrons at energy $E$ with spin $\sigma$ coupled
with the holes of the same energy but opposite spin $-\sigma$;
$i=(i_a,i_b)$ are the lattice indices in the coordinate system
defined by the crystalline $a$ and $b$ axes (Hereafter the lattice
constant is chosen to be unity); the single particle Hamiltonian
${\mathcal H}_{ij}=-t\delta_{i+\tau,j}+
W_{ij}+(U_{i}+S_{i,-\sigma}) \delta_{ij}$, where $t$ is the
nearest-neighbor hopping integral and $\tau=(\tau_a,\tau_b)=(\pm
1,0)$ or $(0,\pm 1)$ represent the relative position of those
sites nearest neighboring to the $i$th site;
$W_{ij}=(-1)^{i_a+i_b}W_{0} \delta_{ij}$ for the conventional CDW
ordering while $W_{ij}=(-1)^{i_a+i_b}
iW_{d}\mbox{sgn}(\tau_{a}^{2}-\tau_{b}^{2})$ for the DDW ordering,
$S_{i,\sigma}=\sigma (-1)^{i_a+i_b} S_0$ represents the SDW order
parameter; $U_{i}$ is the scattering potential to model the
impurities, surface or interface. The quantity
$\Delta_{ij}=\Delta_{s}\delta_{ij}$ for the $s$-wave
superconducting order parameter while $\Delta_{ij}=\Delta_{d}
\mbox{sgn}(x_{ij}^{2}-y_{ij}^{2})$ for the DSC order parameter. We
note that the CDW, SDW, and DDW fields participate in the diagonal
elements of the $2\times 2$ BdG matrix equations in spin space
while the superconducting order parameter appears as off-diagonal
elements. Of most interest to us in the present work, as explained
above, we consider the tunnel junction comprising a normal metal
on the left hand side and a superconductor on right hand side with
a [110] oriented surface. We choose that the interface runs along
the $y$ direction while the normal to the interface is parallel to
the $x$ direction, along which the transport takes place. The
insulating barrier at the interface is modeled by a scattering
potential, $U_{i}=U_{0}\delta_{i_x,0}$. In addition, we assume
that the order parameter is identically zero in the normal metal
and constant in the superconductor. In general, due to the
proximity effect, the order parameter is degraded when the
interface is approached from the interface within a characteristic
length. However, the discontinuity of the pair field remains at
the interface and the physics discussed here is not changed
qualitatively. By using the translation invariance of the system
along the specular interface with spacing $\sqrt{2}$, the
differential conductance (in units of the conductance quantum
$e^2/h$) is calculated within the framework of the
Blonder-Tinkham-Klapwijk (BTK)~\cite{BTK82} theory:
\begin{equation}
G(E)=\frac{1}{N_y}\sum_{k_{y},\sigma} [1-\vert
r_{n,\sigma}(E,k_y)\vert^{2} +\vert r_{a,\sigma}(E,k_y)\vert^{2}]
\end{equation}
where $Ny \sqrt{2}$ is the linear dimension of the junction along
the $y$ direction, $r_{n}$ and $r_{a}$ are respectively the normal
and Andreev reflection amplitude for a beam of electrons with spin
$\sigma$ incident from the normal  metal at a fixed energy $E$ and
the transverse wavevector $k_y=\sqrt{2}\pi l_y/N_y$ with
$l_y=0,\dots,N_y-1$. We determine the values of $r_{n}$ and
$r_{a}$ by matching the wavefunction. Away from the interface, the
solution to the BdG equations consists of the incoming electron,
reflected and transmitted electron and hole waves. The normalized
zero-temperature tunneling conductance is defined as:
$g(E)=G(E)/G_{nn}(E)$, where
$G_{nn}(E)=\frac{1}{N_y}\sum_{k_y,\sigma}
\frac{4}{4+U_{0}^{2}/t_{\perp}^{2}}$ with $t_{\perp}=-2t\cos
(k_{y}/\sqrt{2})$ is the tunneling conductance through a normal
junction of the identical interface condition. We remark that the
BTK formula can be considered as an extended version of the
Landauer-B\"{u}ttiker~\cite{Landauer70} formula for normal
metal/superconductor junctions. The latter has found spectacular
success in the description of quantum coherent transport in
mesoscopic normal conductors. At the same time, we exactly
diagonalize Eq.~(\ref{EQ:BdG}) to calculate the LDOS for a
[110]-oriented surface of the superconductor, which is defined as:
\begin{equation}
\rho_{i_x}(E) =\frac{1}{N_y}\sum_{n,k_y}[\vert
u_{i_x,k_y}^{n}\vert^{2}\delta(E-E_{k_y}^{n}) +\vert
v_{i_x,k_y}^{n}\vert^{2} \delta(E+E_{k_y}^{n})]\;,
\end{equation}
where for a fixed value of $k_y$, the summation is over the
complete set of eigenstates.

\begin{figure}[t]
\centerline{\psfig{figure=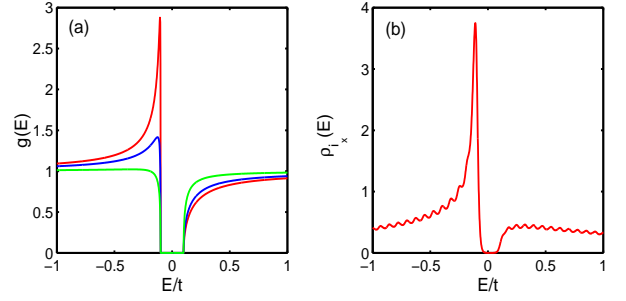,height=4cm,width=8cm,angle=0}}
\caption{The normalized tunneling conductance (a) of a
[110]-oriented N/CDW junction for various strength of the
tunneling barrier, $U_{0}=10t$ (red line), $4t$ (blue line), and
$1t$ (green line). Also shown in (b) the LDOS near the
[110]-oriented surface. The amplitude of the CDW, $W_{0}=0.1t$.}
\label{FIG:CDW}
\end{figure}

{\em Tunneling into CDW.} In the case of normal-metal/CDW
junction, the reflection amplitudes are obtained as:
\begin{subequations}
\begin{eqnarray}
r_{n,\sigma}&=&\frac{-\bigl(\frac{U_{0}}{t_{\perp}}+\frac{i(E-W_0)}
{\sqrt{E^{2}-W_{0}^{2}}}\bigr)+i\mbox{sgn}(E)}
{\bigl(\frac{U_{0}}{t_{\perp}}+\frac{i(E-W_0)}
{\sqrt{E^{2}-W_{0}^{2}}}\bigr)+i\mbox{sgn}(E)}\;,\\
r_{a,\sigma}&=&0\;. \label{EQ:CDW}
\end{eqnarray}
\end{subequations}
Notice that for the CDW, SDW, and DDW orderings (as considered
below), the electron correlation is through the particle-hole
channel so that the charge is conserved and the Andreev reflection
process is absent. In Fig.~\ref{FIG:CDW}(a), we show the tunneling
conductance $g(E)$ as a function of energy for various strength of
the tunneling barrier. The surface LDOS $\rho(E)$ is displayed in
Fig.~\ref{FIG:CDW}(b). As shown in Fig.~\ref{FIG:CDW}, both the
tunnel conductance and the LDOS vanish when the energy is below
the energy gap $W_{0}$.  As the tunnel barrier becomes more
opaque, the profile of the energy dependence of the tunneling
conductance resembles that of the surface LDOS. Both $g(E)$ and
$\rho(E)$ differ dramatically from those of a
normal-metal/$s$-wave superconductor junction~\cite{BTK82}, by
exhibiting no coherent peak at one of the gap edges. They are also
asymmetric with respect to the Fermi energy $E=0$. We also find
the property that $g(E,W_0)=g(-E,-W_0)$ and
$\rho(E,W_0)=\rho(-E,-W_0)$, which show the strong dependence of
both the tunneling conductance and LDOS on the macroscopic phase
of the CDW.

\begin{figure}[t]
\centerline{\psfig{figure=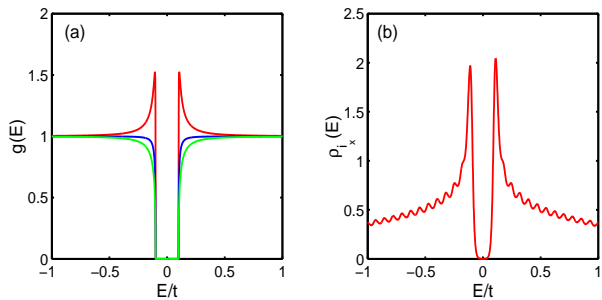,height=4cm,width=8cm,angle=0}}
\caption{The normalized tunneling conductance (a) of a
[110]-oriented N/SDW junction for various strength of the
tunneling barrier, $U_{0}=10t$ (red line), $4t$ (blue line), and
$1t$ (green line).  Also shown in (b) the LDOS near the
[110]-oriented surface. The amplitude of the SDW, $S_{0}=0.1t$.}
\label{FIG:SDW}
\end{figure}

{\em Tunneling into SDW.} In the case of normal-metal/SDW
junction, we arrive at the reflection amplitudes as:
\begin{subequations}
\begin{eqnarray}
r_{n,\sigma}&=&\frac{-\bigl(\frac{U_{0}}{t_{\perp}}+\frac{i(E-\sigma
S_0)} {\sqrt{E^{2}-S_{0}^{2}}}\bigr)+i\mbox{sgn}(E)}
{\bigl(\frac{U_{0}}{t_{\perp}}+\frac{i(E-\sigma S_0)}
{\sqrt{E^{2}-S_{0}^{2}}}\bigr)+i\mbox{sgn}(E)}\;,\\
r_{a,\sigma}&=&0\;. \label{EQ:SDW}
\end{eqnarray}
\end{subequations}
In contrary to the CDW case, where the spin is degenerate for the
tunnel conductance, the reflection amplitude for the SDW is spin
dependent. In particular, there holds the following symmetry
property of the tunneling conductance between different spin
bands: $g_{\sigma}(E,S_0)=g_{-\sigma}(-E,S_0)$. It follows
immediately that the total tunneling conductance is symmetric with
respect to the Fermi energy, i.e., $g(E,S_0)=g(-E,S_0)$. As shown
in Fig.~\ref{FIG:SDW}, both the tunneling conductance and the LDOS
vanishes when the energy below the energy gap $S_0$. The profile
of $g(E)$ also resembles to that of $\rho(E)$ as the strength of
the tunnel barrier is increased. In addition, the coherent peak
shows up at the gap edge $E=\pm S_0$ in both $g(E)$ and $\rho(E)$.
Therefore, the overall behavior of the tunneling conductance of
the N/SDW junction is qualitatively similar to the
normal-metal/$s$-wave superconductor junction.

\begin{figure}
\centerline{\psfig{figure=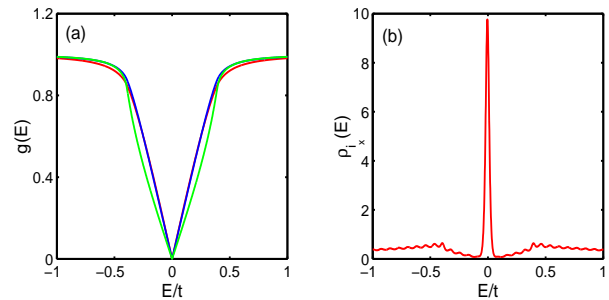,height=4cm,width=8cm,angle=0}}
\caption{The normalized tunneling conductance (a) of a
[110]-oriented N/DDW junction for various strength of the
tunneling barrier, $U_{0}=10t$ (red line), $4t$ (blue line), and
$1t$ (green line). Also shown in (b) the LDOS near the
[110]-oriented surface. The amplitude of the DDW, $W_{d}=0.1t$.}
\label{FIG:DDW}
\end{figure}

{\em Tunneling into DDW.} In the case of normal-metal/DDW
junction, the reflection amplitudes are obtained as:
\begin{subequations}
\begin{eqnarray}
r_{n,\sigma}&=&\frac{-\bigl(\frac{U_{0}}{t_{\perp}}+\frac{iE}
{\sqrt{E^{2}-W_{\mathbf{k}}^{2}}+W_{\mathbf{k}}}\bigr)+i\mbox{sgn}(E)}
{\bigl(\frac{U_{0}}{t_{\perp}}+\frac{iE}
{\sqrt{E^{2}-W_{\mathbf{k}}^{2}}+W_{\mathbf{k}}}\bigr)+i\mbox{sgn}(E)}\;,\\
r_{a,\sigma}&=&0\;, \label{EQ:DDW}
\end{eqnarray}
\end{subequations}
where $W_{\mathbf{k}}=-4W_d \sin(k_y/\sqrt{2})$. The energy gap
for both the CDW and SDW ordering, as discussed above, is constant
on the Fermi surface in the momentum space. The bulk energy gap of
the DDW ordering is momentum-dependent, and of the form $\cos k_x
-\cos k_y$, which shows clearly that the DDW energy is closed
along the nodal directions on the Fermi surface. In
Fig.~\ref{FIG:DDW}, we show the tunneling conductance of the N/DDW
junction and the surface LDOS of the superconductor at the DDW
state. Noticeably, a sharp peak appears at $E=0$ in the surface
LDOS, which indicates the existence of the midgap states at the
[110]-oriented surface of the DDW ordering. However, the tunnel
conductance exhibits a V-shaped feature, which persists even when
the tunnel barrier is opaque. The V-shape originates from the
momentum dependence of the energy gap of the DDW ordering. In
particular, the tunneling conductance does not show a zero-bias
anomaly, it instead dips to zero. It can also be seen clearly that
when $E=0$, $\vert r_{n} \vert^{2}=1$, which gives $g(E)=0$,
regardless of the strength of the tunnel barrier. Therefore, in
the tunneling limit, the profile of the conductance is strikingly
different from the surface LDOS. Physically,  the energy gap of
the DDW ordering comes from the particle-hole pairs. In contrast
to the superconducting state, there exists no Andreev reflection
process to convert, at the interface, the single particle current
into the supercurrent through the condensate, so that all
quasiparticles with energy below the momentum-dependent DDW gap
are completely reflected. Consequently, although there exist
midgap states at the [110]-oriented surface of the superconductor
in the DDW ordering, these states do not carry charge. However, as
in the DSC case~\cite{Zhu00}, the midgap states in the DDW state
can also lead to a giant magnetic moment when a strong in-plane
magnetic field is applied, the study of which is beyond the scope
of current work. At this stage, we can conclude that no midgap
states exist at the surface of the superconductor with either CDW
or SDW ordering, but they really show up in the DDW ordering.

\begin{figure}
\centerline{\psfig{figure=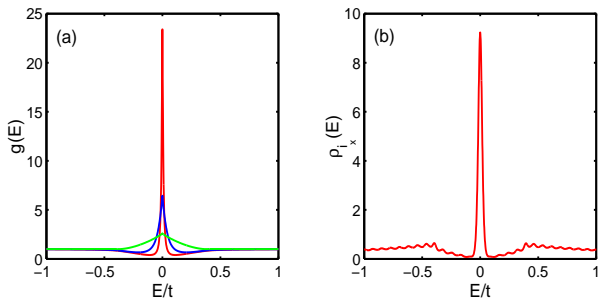,height=4cm,width=8cm,angle=0}}
\caption{The normalized tunneling conductance (a) of a
[110]-oriented N/DSC junction for various strength of the
tunneling barrier, $U_{0}=10t$ (red line), $4t$ (blue line), and
$1t$ (green line). Also shown in (b) the LDOS near the
[110]-oriented surface. The amplitude of the DSC,
$\Delta_{d}=0.1t$.} \label{FIG:DSC}
\end{figure}

{\em Tunneling into DSC.} The tunneling conductance of
normal-metal/DSC junctions, has been studied very intensively in
the continuum model. Considering that the Andreev reflection
process should take place even above $T_{c}$ in the preformed pair
scenario, or in the spin-charge separation picture (where the
spinons form $d$-wave resonant-valence-bond state), we believe it
is still helpful to present, within the lattice model, the results
for the DSC case. We arrive at the reflection amplitudes:
\begin{subequations}
\begin{eqnarray}
r_{n,\sigma}&=&\frac{\bigl(-2i\frac{U_0}{t_{\perp}}
+\frac{U_{0}^{2}}{t_{\perp}^{2}}\bigr)\vert
E\vert}{-2(\vert E\vert +\sqrt{E^2-\vert
\Delta_{\mathbf{k}}\vert^{2}})+\frac{U_{0}^{2}}{t_{\perp}^{2}}\vert
E\vert}\;,\\
r_{a,\sigma}&=&\frac{2\mbox{sgn}(E) \vert
\Delta_{\mathbf{k}}\vert}{-2(\vert E\vert +\sqrt{E^2-\vert
\Delta_{\mathbf{k}}\vert^{2}})+\frac{U_{0}^{2}}{t_{\perp}^{2}}\vert
E\vert}\;,
\end{eqnarray}
\label{EQ:DSC}
\end{subequations}
where $\Delta_{\mathbf{k}}=-4\Delta_{d}\sin (k_y/\sqrt{2})$. As
shown in Fig.~\ref{FIG:DSC}, the conductance at zero energy does
not vanish. In particular,  when the strength of the tunneling
barrier is increased, a sharp peak occurs at zero energy in the
conductance. In the opaque limit of the tunneling barrier, the
profile of the conductance resembles that of the surface LDOS,
which also exhibits a zero-energy peak, indicating the existence
of midgap states. As we have emphasized above, the superconducting
energy gap originates from the formation of Cooper pairs due to
the electronic correlation in the particle-particle channel.
Therefore, the charge is not conserved in a superconductor, and
the Andreev reflection can happen in the interface between the
normal metal and superconductor. It can be found from
Eq.~(\ref{EQ:DSC}) that, at $E=0$, $r_{n,\sigma}=0$ and
$r_{a,\sigma}=1$, regardless of the strength of the tunneling
barrier. Therefore, the existence of the midgap states in the
superconducting state corresponds to the perfect Andreev
reflection process. In the phase fluctuation scenario, the
long-range phase coherence is absent but the short-range
correlation is still retained, the localized nature of the midgap
states should be robust against the phase fluctuations. Therefore,
if the phase fluctuation is the mechanism for the PG, we expect
that the zero-bias conductance anomaly is still observable.

In summary, we have studied the tunneling conductance and the
surface LDOS of a [110]-oriented high-$T_c$ superconductor in the
underdoped regime. The distinct features among the different kinds
of ordering---CDW, SDW, DDW, and DSC, are investigated. We have
shown, in the opaque limit of the tunneling barrier, that: (i) for
the CDW ordering, asymmetric tunneling conductance and surface
LDOS with vanishing intensity within the gap; (ii) for the SDW
ordering, symmetric tunneling conductance and surface LDOS with
vanishing intensity within the gap; (iii) for the DDW ordering,
V-shaped tunneling conductance but zero-energy-peaked surface
LDOS; (iv) for the DSC ordering, zero-energy peaked tunneling
conductance and surface LDOS. These different features can serve
to distinguish the competing scenarios for the mechanism of the
PG. We have also found the existence of surface midgap states in
the DDW ordering, which is, to best of our knowledge, the first
example occurring in a gapped normal system. Also in this case,
the correspondence between the tunneling conductance and the
surface LDOS is broken down.

{\bf Acknowledgments}: The author thanks S. A. Trugman for helpful
discussions. This work was supported by the Department of Energy
through the Los Alamos National Laboratory.

\end{document}